# Visualizing Magnetic Order in Self-Assembly of Superparamagnetic Nanoparticles

Xingyuan Lu[1,2†], Ji Zou[1,3†], Minh Pham[1,4], Arjun Rana[1], Chen-Ting Liao[5], Emma Cating Subramanian[5], Xuefei Wu[6], Yuan Hung Lo, Charles S. Bevis[5], Robert M. Karl Jr[5], Serban Lepadatu[7], Young-Sang Yu[8], Yaroslav Tserkovnyak[1], Thomas P. Russell[6], David A. Shapiro[8], Henry C. Kapteyn[5], Margaret M. Murnane[5], Robert Streubel[9] and Jianwei Miao[1*]

[1]*Department of Physics & Astronomy and California NanoSystems Institute, University of California, Los Angeles, CA 90095, USA and STROBE Science and Technology Center*

[2]*School of Physical Science and Technology, Soochow University, Suzhou 215006, China*

[3]*Department of Physics, University of Basel, Klingelbergstrasse 82, CH-4056 Basel, Switzerland*

[4]*Department of Mathematics, University of California, Los Angeles, CA 90095, USA and STROBE Science and Technology Center*

[5]*JILA and Department of Physics, University of Colorado and NIST, 440 UCB, Boulder, Colorado 80309, USA and STROBE Science and Technology Center*

[6]*Polymer Science and Engineering Department University of Massachusetts, Amherst, Massachusetts 01003, USA*

[7]*Jeremiah Horrocks Institute for Mathematics, Physics and Astronomy, University of Central Lancashire, Preston, PR1 2HE, U.K.*

[8]*Advanced Light Source, Lawrence Berkeley National Laboratory, Berkeley, CA 94720, USA*

[9]*Department of Physics & Astronomy, University of Nebraska-Lincoln, Lincoln, NE 68588, USA*

**We use soft x-ray vector-ptychographic tomography to determine the three-dimensional magnetization field in superparamagnetic nanoparticles self-assembled at the liquid-liquid**

**interface and reveal the magnetic order induced by layered structure. The spins in individual nanoparticles become more aligned with increasing number of layers, resulting in a larger net magnetization. Our experimental results show a magnetic short-range order in the monolayer due to the proliferation of thermally induced magnetic vortices and a magnetic long-range order in the bilayer and trilayer, stemming from the strengthened dipolar interactions that effectively suppress thermal fluctuations. We also observe a screening effect of magnetic vortices and the attractive interaction between the magnetic vortices with opposite topological charges. Our work demonstrates the crucial role of layered structure in shaping the magnetization of nanoparticle assemblies, providing new opportunities to modulate these properties through strategic layer engineering.**Ferromagnetic liquid droplets, generated by the mechanical jamming of superparamagnetic nanoparticles at liquid-liquid interfaces, introduce a novel paradigm in magnetic materials due to their combined mechanical and magnetic adaptability [1-3]. These magnetostatically interconnected nanostructures influence magnetic field-driven actions, such as translation, rotation, and actuation, holding great promise in diverse fields like data storage, magnetic sensing, and microrobotics [4-6]. The burgeoning interest in their broad application potential has propelled numerous experimental investigations, aiming to deepen our understanding of their magnetic properties and, in turn, optimize their structural and magnetic order for better utility [7-19]. Much research has delved into both the macroscopic and nanoscopic characteristics of nanoparticle assemblies, including the thickness-dependent, dipole-induced ferromagnetic order in three-dimensional (3D) crystalline nanoparticle assemblies [20-22] and its persistence in 1D and quasi-2D nanostructures that lack crystalline symmetry [23,24]. However, despite these

advances, accessing the magnetization configuration within individual nanoparticles and deciphering the formation of microscopic magnetization patterns in structurally disordered layered nanoparticle assemblies have remained elusive. This is largely due to the difficulty to access the 3D magnetic moments of the nanoparticle system without assuming any spatial symmetry.

The recent development of soft x-ray vector-ptychographic tomography holds promise as a solution to this challenge [25,26]. Rooted in the principle of coherent diffractive imaging [27-29], this method offers a high spatial resolution with a sensitivity that is about two orders of magnitude higher than the hard x-ray counterpart due to its enhanced scattering cross section [30,31]. In this Letter, we use soft x-ray vector-ptychographic tomography to image the 3D magnetization field of individual nanoparticles and quantify the influence of layered structure on the magnetic order within densely packed assemblies. We observed that magnetic vortices proliferate in the monolayer subject to thermal fluctuations that destroys spin correlations and collinear magnetic ordering. In contrast, fewer vortices are observed in the bilayer and trilayer. This phenomenon is attributed to the enhanced magnetic dipole interactions present in nanoparticle multilayers that is four times stronger than that in the monolayer. Our findings highlight the significant impact of layered structure on the magnetic ordering in nanoparticle assemblies, opening up avenues for tailored modulation of magnetic properties through targeted layer design.

Our experiment used superparamagnetic $Fe_3O_4$ nanoparticles with magnetic cores about 22 nm [1-3]. The sample was prepared by drop-casting aqueous nanoparticles dispersions onto silicon nitride membranes that were dried under ambient conditions (Supplemental Material [32]). The tomographic imaging was performed at the COSMIC imaging beam line at the Advanced Light Source, where circularly polarized x-rays with an energy tuned to the Fe $L_3$ edge (713 eV) were focused onto the sample by a Fresnel zone plate (Fig. 1). To achieve 3D vector reconstruction, two tilt series of data were acquired from the sample at two in-plane rotation angles (0° and 120°), respectively. At each in-plane angle, the sample was rotated around the x-axis to produce a tilt series of diffraction patterns. At each tilt angle, the focused beam was raster-scanned across the sample and two sets of diffraction patterns were collected with left- and right-circularly polarized x-rays (Table S1 in Supplemental Material [32]). The complete data contains a total of 213,996 diffraction patterns, from which 70 images were reconstructed by an iterative phase retrieval algorithm [33]. After normalization, background subtraction, and image alignment using the center of mass method [34], each pair of the oppositely polarized images was summed to produce 35 projections for 3D scalar reconstruction. The scalar reconstruction was performed with Real Space Iterative Reconstruction (RESIRE), a powerful tomographic algorithm that enables to reconstruct 3D object from a limited number of projections with automated angular refinement [35,36] (Supplemental Material [32]). Figure 2a-d shows the 3D scalar reconstruction of the nanoparticle assembly (grey regions), where the nanoparticles form monolayers, bilayers, and trilayers. The 3D coordinates of the nanoparticles were traced from the reconstruction and used to calculate the radial distribution function, which exhibits a structural short-range order with the first maximum at 32.5 nm (Fig. 2e). The difference between the left- and right-circularly polarized images, taken at otherwise same conditions, produce 35 projections used for the 3D vector reconstruction. The 3D magnetization vector field of the sample was reconstructed using a vector tomography method described elsewhere [25,26] and a 3D support derived from the scalar reconstruction (Supplemental Material [32]). As the sample has layered structure with a maximum of three layers, the 35 projections were sufficient to produce good 3D scalar and vector reconstructions (Supplemental Material [32]).

Quantitative analysis of the magnetic order in the superparamagnetic nanoparticle assembly was performed in terms of the net magnetization inside each nanoparticle, referred to as a macrospin from the 3D vector reconstruction (Fig. 2a). Figure 2b-d shows the distribution of the macrospins in the monolayer, bilayer, and trilayer unveiling macrospins that increase with the number of layers. These variations are due to thermal spin fluctuations at room temperature since the saturation magnetization of each nanoparticle is identical. Figure 2f and g show the histograms of the magnitude of the macrospin with the angle defined by $\cos\theta = \boldsymbol{S} \cdot \overline{\boldsymbol{S}}$, where $\boldsymbol{S}$ is normalized macrospin and $\overline{\boldsymbol{S}}$ is the normalized mean macrospin in each region. The mean and standard deviations of the macrospin magnitudes in the three regions is 0.17±0.1 (monolayer), 0.25±0.12 (bilayer), and 0.33±0.15 (trilayer), respectively, and the macrospin angle is better aligned with increasing number of layers. The formation of magnetic vortices and anti-vortices in the nanoparticle assembly, which plays a crucial role in determining magnetic order of a system [37], was quantified in terms of the vortex density [38]:

$$\rho(x,y) = \frac{1}{\pi}\hat{\boldsymbol{z}} \cdot (\partial_x \boldsymbol{S} \times \partial_y \boldsymbol{S}). \quad (1)$$

The total vortex number within an area $\Omega$ is given by:

$$Q = \int_\Omega \rho \, dxdy = \frac{1}{2\pi} \int_{\partial\Omega} \boldsymbol{S}_{\|}^2 \, \nabla\varphi \cdot d\boldsymbol{l}, \quad (2)$$

where $\boldsymbol{S}_{\|} = (S^x, S^y)$ is the planar projection of the vector field $\boldsymbol{S}$, $\varphi$ is its polar angle relative to the $x$ axis, and $\partial\Omega$ denotes the boundary of the area parameterized by $d\boldsymbol{l}$. This reveals the bulk-edge correspondence of magnetic vortices: the total vortex charge inside a region is entirely determined by the magnetic states on its boundary.

For our purpose, we discretize Eq. (1) to determine the vortex number for a triangular region defined by three vertices:

$$\rho = \frac{\hat{\mathbf{z}} \cdot \sum_{<l,l'>} (\mathbf{S}_l \times \mathbf{S}_{l'})}{2\pi}, \qquad (3)$$

where $\hat{\mathbf{z}}$ represents the normal vector of the triangular plane, $\mathbf{S}$ is the normalized macrospin sitting on the vertices, and the sum runs over the three edges ($l$ is the vertex adjacent to $l'$ in the counterclockwise direction). Given that the triangle serves as the foundational building block for arbitrary 2D lattices including disordered structures, Eq. (3) is applicable to our nanoparticle assembly characterized by its short-range structural nature [39,40]. The total vortex charge within a region can be determined by summing the vortices associated with all the minimal triangles contained within. This summation vanishes in the bulk, leaving only the boundary terms, which is a discrete version of the bulk-edge correspondence of vortices [39,40]. It should be noted that the vortex number, defined by Eq. (3), is not integer-valued with its maximum being $3\sqrt{3}/4\pi$ for normalized macrospins [32]. Figure 3a shows the distribution of magnetic vortices and anti-vortices with $|\rho| \geq 0.25$, whereas a more detailed vortex density map of the system is provided in Fig. S2a. Figure 3b-d illustrates the typical configuration of two representative vortices, a collinear arrangement and two anti-vortices, respectively. We observed that magnetic vortices and anti-vortices persist dominantly in the monolayer (A1 in Fig. 3a) and at the boundary between different layers (A3 in Fig. 3a) where the magnetic dipole interaction is weak and/or spin frustration is enhanced. The occurrence is suppressed in trilayers (A2 in Fig. 3a). This observation was quantified by a larger vortex and anti-vortex density (i.e., $|\rho| \geq 0.25$) in the monolayer than in the bilayer and trilayer (Fig. 3e). By analyzing the nearest-neighbor distances of vortex-antivortex (Fig. 3f), vortex-vortex (Fig. 3g), and antivortex-antivortex pairs (Fig. 3h), we observed a tendency for vortices and antivortices to attract to each other. The distance between vortex-antivortex pairs was determined to be 22.3±1.4 nm, fitting a generalized extreme value distribution that accounts for the asymmetry in the observed distance distribution (Fig. 3f). In contrast, the vortex-vortex (Fig. 3g) and antivortex-antivortex pairs (Fig. 3h) were found to stabilize at comparatively larger distances of 43.9±3.9 nm and 58.2±4.9 nm, respectively. Such an attractive interaction between magnetic vortices of opposite topological charges leads to the vortex screening effect, as discussed later.

The magnetic order in different regions was further examined by the spin-spin correlation function, defined as $C_{spin}(|\boldsymbol{R}|) = \langle \boldsymbol{S}(\boldsymbol{r})\boldsymbol{S}(\boldsymbol{r}+\boldsymbol{R})\rangle_r$ with $|\boldsymbol{R}|$ as the pair distance, which measures the degree of collinearity of the spins. The blue curve in Figure 3i presents the spin-spin correlation function in the monolayer, exhibiting an exponential decay to a small value of approximately 0.19 at large distances, with a spin correlation length of $\xi \approx 20$ nm. This suggests a low degree of collinearity among spins in the monolayer. This spin correlation length is comparable to the size of the superparamagnetic nanoparticles, corroborating short-range ordered spins and absent spin correlation even between two nearest macrospins (Fig. 2e). In contrast, the spin correlation functions for the bilayer and trilayer, as shown by the green and orange curves in Fig. 3i, saturate at sizable values of 0.4 and 0.5 at large distances, $|\boldsymbol{R}| > 500$ nm. This behavior indicates the presence of the magnetic long-range order and implies a higher degree of spin collinearity within the bilayer and trilayer structures. This distinction is attributed to the different coupling strength of the magnetic dipole forces in the nanoparticle assembly, which was predicted to stabilize the 2D magnetic order [41]. Based on the averaged magnetic moments of closely packed nanoparticles from the vector reconstruction (1:1.48:1.91), the dipole energy ratio is 1:2.2:3.6 for the monolayer, bilayer and trilayer. While thermal fluctuations lead to the proliferation of magnetic vortices and anti-vortices in the monolayer due to the relatively weak dipolar interaction, they are effectively suppressed in the bilayer and trilayer because of enhanced dipolar coupling, facilitating the emergence of the magnetic order.

We further confirmed these results by evaluating the vortex correlation function defined as, $C_{vortex}(|\mathbf{R}|) = \langle \rho(\mathbf{r})\rho(\mathbf{r}+\mathbf{R})\rangle_r$, where the vortex charge $\rho$ is defined in Eq. (3). Figure 3j shows the vortex correlation functions for the monolayer (blue), bilayer (green), and trilayer (orange). At $|\mathbf{R}| = 0$, magnetic vortices exhibit positive correlation, which essentially reflects the average squared density of the vortices. A higher vortex density was observed in the monolayer with $C_{vortex}(0) = 7.2 \times 10^{-3}$, which destroys the long-range magnetic order, in contrast to the comparatively lower values in the bilayer ($4 \times 10^{-3}$) and trilayer ($2.9 \times 10^{-3}$). Interestingly, we also found that the vortex correlation functions decrease to negative values at short distances before decaying to zero as shown in Fig. 3j, indicative of a screening effect among vortices. Magnetic vortices are surrounded by others with opposite charges, a result of their mutual attractive interactions. This leads to a negative vortex correlation at short range, while simultaneously screening the vortices from distant ones, resulting in zero correlation at larger distances. The vortex correlation in the monolayer shows greater fluctuations around zero at large distances than in the bilayer and trilayer, primarily due to noise in vortex positioning. The bilayer and trilayer, having fewer vortices, experience inherently lower noise contributions to their correlation functions.

The reconstructed magnetization vector field and magnetic correlations were compared with magnetic simulations to validate the correctness of reconstruction in a physical context and establish the link between structural short-range order and magnetic ordering. We modeled the superparamagnetic nanoparticle assembly using the experimentally determined 3D coordinates as direct input at 300 K by means of micromagnetic Monte Carlo simulations with Boris Spintronics (Supplemental Material [32]) [42,43]. Each nanoparticle is approximated as a macrospin with a variable magnetization length that interacts with adjacent nanoparticles through the dipolar magnetic field. The resulting magnetic hysteresis loop reveals a near-zero remanence value, typical for closely packed superparamagnetic nanoparticles (Fig. 4a). The magnetization length averaged over the generated ensemble follows the expected Maxwell-Boltzmann distribution, with the mean value approximately equal to the saturation magnetization [42,44]. By averaging the magnetization first over the generated ensemble and calculating the probability distribution, we obtained an averaged magnetization length close to zero, confirming the superparamagnetic origin of the near-zero remanence (Fig. 4a). Quantifying the short-range order of the entire macrospin system in terms of the spin-spin correlation function (Fig. 4b) and the vortex-vortex correlation function (Fig. 4c) reveals a trend in good agreement with the experimental results. This pertains, in particular, to the distribution and histogram of the vortex density which reveal, for both experimental (Figs. 3e and S2a) and modeled (Fig. S2b and S3) data, a larger vortex and anti-vortex density in the monolayer than in the bilayer and trilayer. Additionally, more magnetic vortices and anti-vortices form at the boundary of different layers (Fig. 4d), which corroborates the dipolar origin of the spin frustration giving rise to the vortex configuration.

In conclusion, soft x-ray vector-ptychographic tomography revealed the room-temperature 3D magnetization vector field and magnetic order in self-assembled superparamagnetic nanoparticles. Our results confirm on the nanometer scale that the thermal stability of the magnetization within each nanoparticle and the magnetization configuration depend on layered structure and the corresponding strength of the dipolar interaction. With increasing of number of layers, the net magnetization of each superparamagnetic nanoparticle increases due to suppressed spin fluctuation. Quantitative analysis of the spin-spin and vortex-vortex correlation functions revealed a long-range macrospin order in the bilayer and trilayer and the formation of magnetic vortices and anti-vortices in the monolayer and at the boundary between different layers. We measured the distance between magnetic vortex and antivortex pairs was measured to be 22.3±1.4 nm, while the magnetic vortex-vortex and antivortex-antivortex pairs are stabilized at comparatively larger distances of 43.9±3.9 nm and 58.2±4.9 nm, respectively. These experimental results were further validated by micromagnetic Monte Carlo simulations based on the experimental 3D coordinates of the superparamagnetic nanoparticles. Our work elucidates the previously reported stable remanent magnetization in ferromagnetic liquid droplets [1-3], where the formation of multilayers of superparamagnetic nanoparticles enhanced the magnetic order and spin correlation (ferromagnetism). Our experiment demonstrates the key role of layered structures in determining the magnetic properties of nanoparticle assemblies. This indicates that layered structures offer a novel degree of freedom to tailor the magnetic properties of superparamagnetic assemblies, which is crucial for a variety of applications, ranging from data storage to microrobotics and biomedicine.

This work was primarily supported by STROBE: a National Science Foundation Science and Technology Center under award DMR1548924. J.M. acknowledges partial support by the U.S. Air Force Office Multidisciplinary University Research Initiative (MURI) program under award no. FA9550-23-1-0281. R.S. acknowledges support by the National Science Foundation, Division of Materials Research under grant no. 2203933. X.Y.L. acknowledges the support by the National Natural Science Foundation of China (No. 12204340). Soft x-ray vector-ptychographic tomography experiments were performed at COSMIC and used resources of the Advanced Light Source, which is a US Department of Energy Office of Science User Facility under contract number DE-AC02- 05CH11231. X. W. and T.P.R. were supported by the U.S. Department of Energy, Office of Science, Office of Basic Energy Sciences, Materials Sciences and Engineering Division under Contract No. DE-AC02-05-CH11231 within the Adaptive Interfacial Assemblies Towards Structuring Liquids program (KCTR16).

†These authors contributed equally to this work.

*Email: j.miao@ucla.edu

**Figures and Figure Captions**

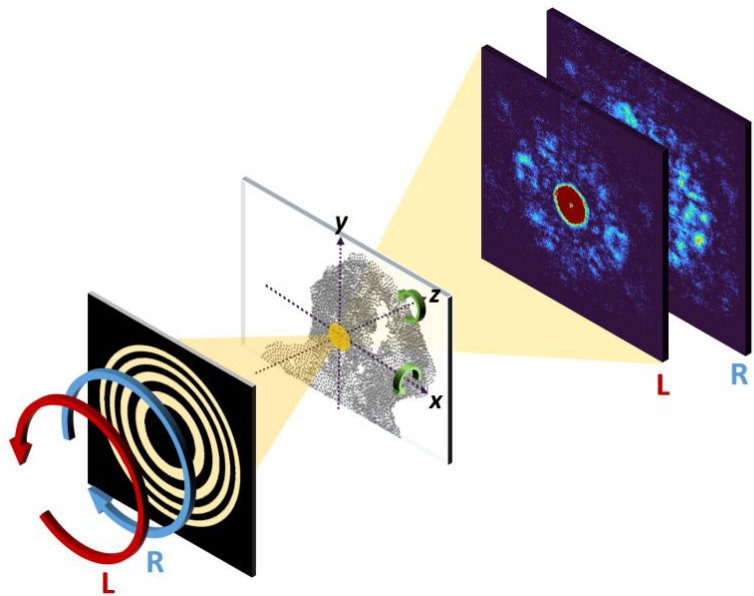

**Fig. 1.** Schematic of soft x-ray vector-ptychographic tomography. X-rays at a photon energy 713 eV, resonant to the Fe $L_3$ edge, were focused by a Fresnel zone plate onto a superparamagnetic nanoparticle assembly. The sample was rotated to different angles around the *x* and *z* axes. At each angle, the focused beam raster-scans across the sample and two sets of diffraction patterns with left- and right-circularly polarized x-rays were recorded by a charged-coupled device.

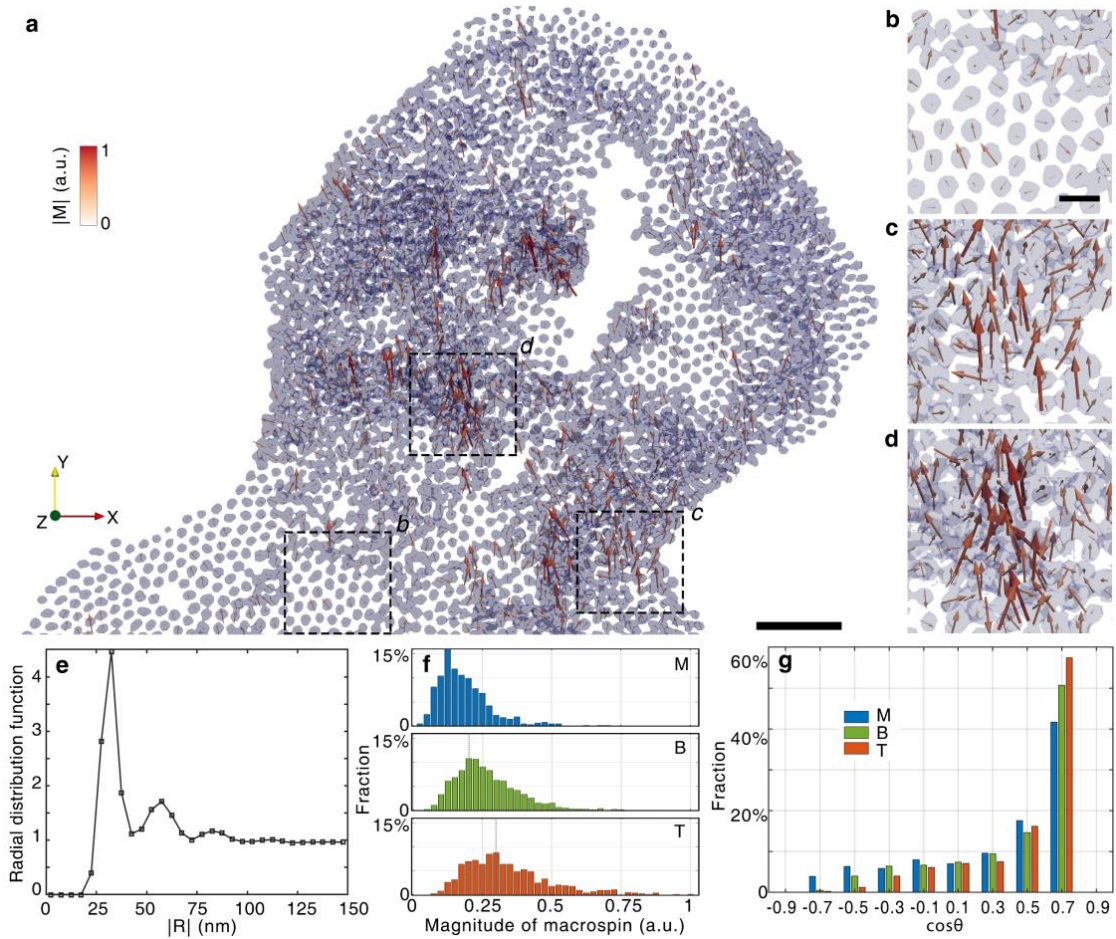

**Fig. 2.** Observation of magnetic order in self-assembly of superparamagnetic nanoparticles. (a) 3D scalar and vector reconstruction of the superparamagnetic nanoparticle assembly, where the 3D magnetization field within each nanoparticle was summed to produce a macrospin (displayed as an arrow). (b-d) Magnified views of the three squared regions in (a), corresponding to a monolayer (b), bilayer (c) and trilayer (d) region, respectively. Colorbar in (a) refers to the macrospin magnitude in (a-d). (e) Radial distribution function calculated from the experimentally determined 3D coordinates of the nanoparticles corroborating structural short-range order. (f) and (g) Histograms of the magnitude and angle of the macrospins in the monolayer, bilayer and trilayer. Scale bar, 200 nm (a) and 50 nm (b).

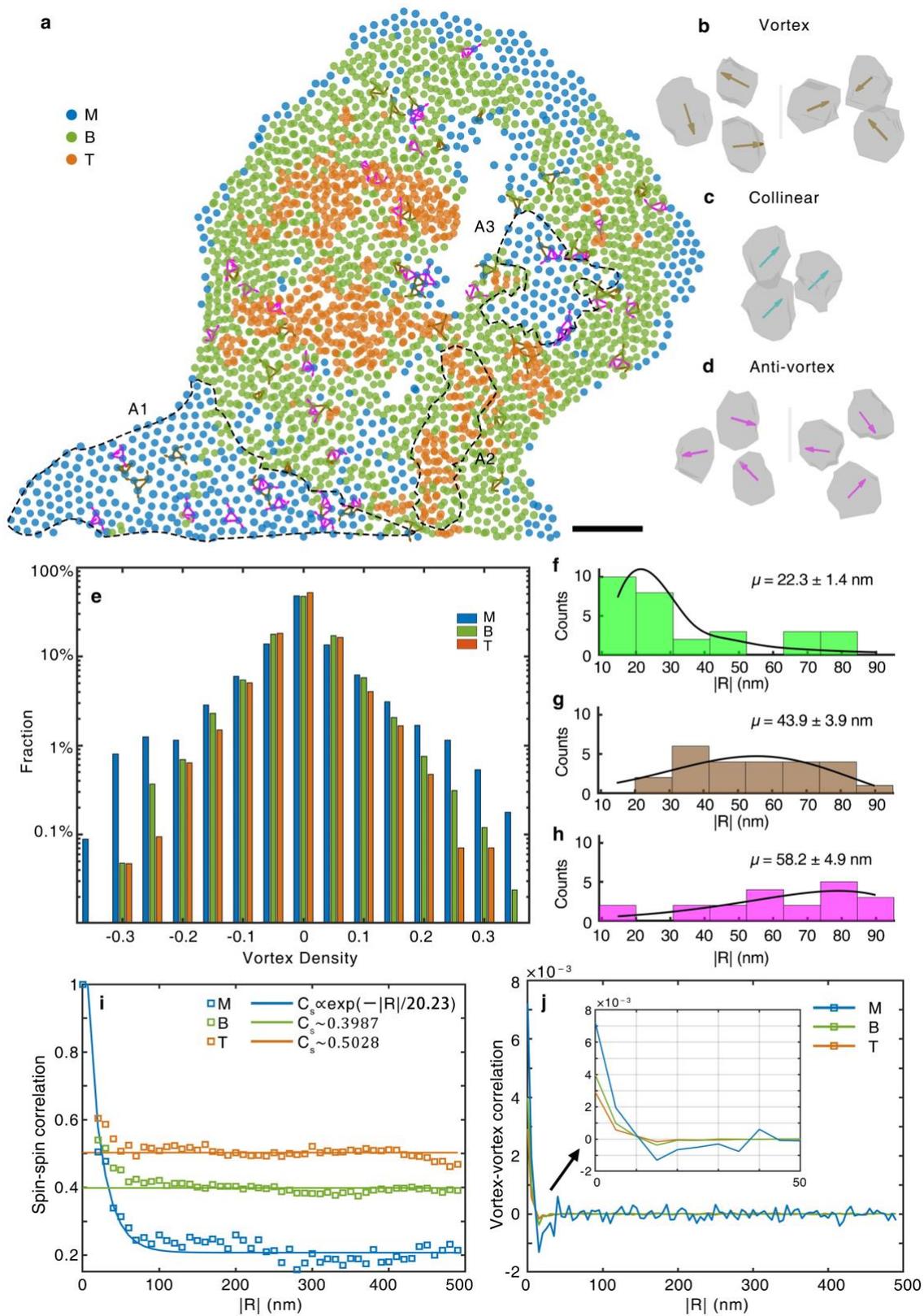

**Fig. 3.** Quantitative characterization of magnetic vortices and anti-vortices in the monolayer, bilayer, and trilayer. (a) Magnetic vortex (brown triangle) and antivortex (pink triangle) distribution in the superparamagnetic nanoparticle assembly, where blue, green, and orange circles represent nanoparticles in the monolayer, bilayer and trilayer regions, respectively. Areas A1, A2 and A3 enclosed by dashed lines highlight a monolayer region full of vortices and anti-vortices, a trilayer region without vortices or anti-vortices, and a boundary of different layers with abundant vortices and anti-vortices, respectively. (b-d) Magnified views of two representative vortices (b), collinear spins (c), and two representative anti-vortices (d). (e) Histogram of magnetic vortex density in the monolayer, bilayer, and trilayer regions, where the total probability of each region is 1. (f-h) Histograms of the nearest-neighbour distances for the vortex and anti-vortex pairs (f), the vortex and vortex pairs (g) and the antivortex and antivortex pairs (h). (i) and (j) Spin-spin correlation function (i) and vortex-vortex correlation function (j) in the monolayer, bilayer, and trilayer. Scale bar in (a), 200 nm.

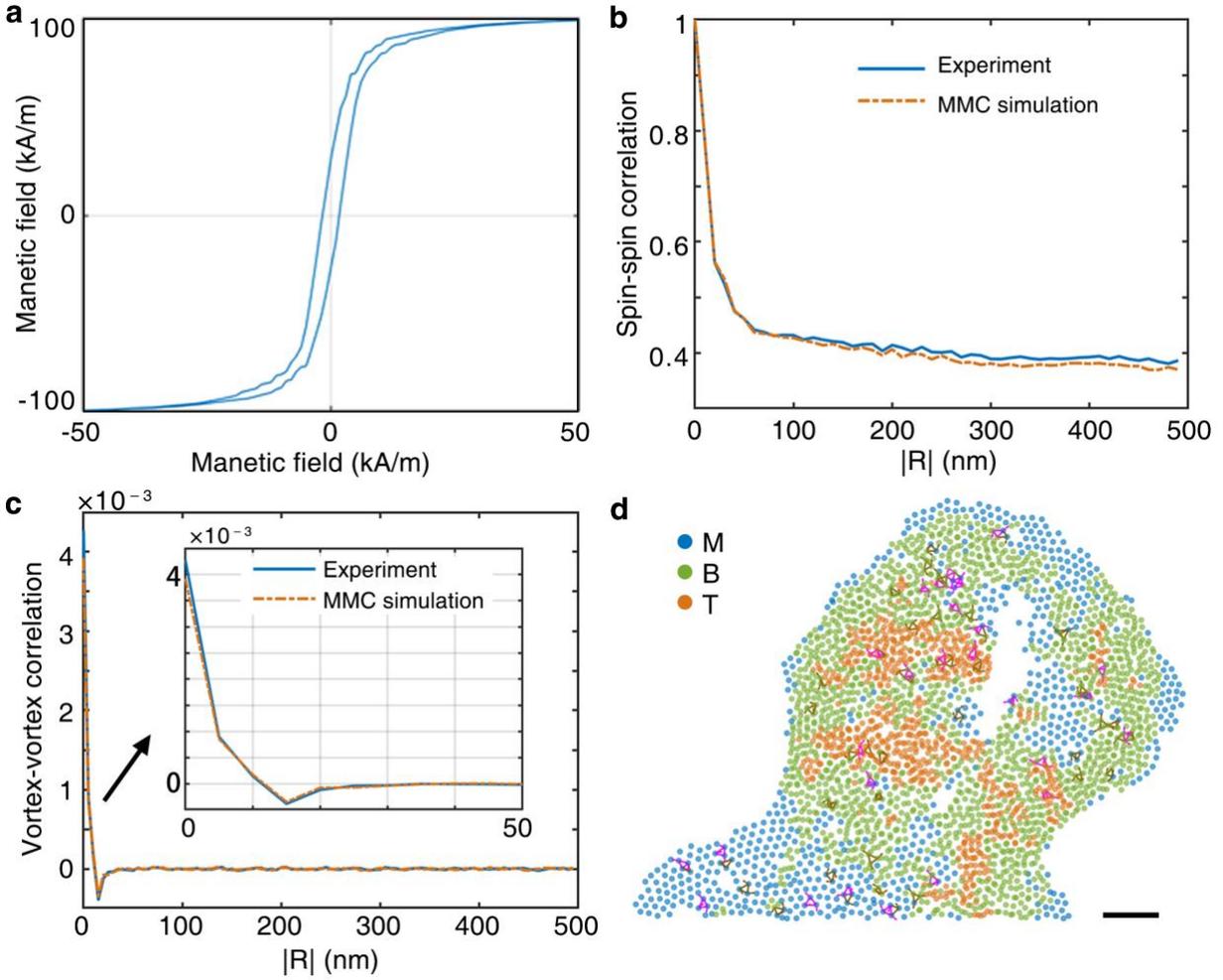

**Fig. 4** Micromagnetic Monte Carlo simulations based on the experimentally determined 3D coordinates and macrospins of the superparamagnetic nanoparticles. (a) In-plane magnetic hysteresis loop with Curie temperature $T_c$ = 850 K revealing weak ferromagnetism. (b) Spin-spin correlation function and (c) vortex-vortex correlation function, revealing a trend in good agreement with the experimental results. (d) Magnetic vortex (brown triangle) and antivortex (pink triangle) distribution, where blue, green, and orange circles represent nanoparticles in the monolayer, bilayer, and trilayer, respectively. Scale bar in (a), 200 nm.

# Supplemental Material for

# Visualizing Magnetic Order in Self-Assembly of Superparamagnetic Nanoparticles

Xingyuan Lu[1,2†], Ji Zou[1,3†], Minh Pham[1,4], Arjun Rana[1], Chen-Ting Liao[5], Emma Cating Subramanian[5], Xuefei Wu[6], Yuan Hung Lo, Charles S. Bevis[5], Robert M. Karl Jr[5], Serban Lepadatu[7], Young-Sang Yu[8], Yaroslav Tserkovnyak[1], Thomas P. Russell[6], David A. Shapiro[8], Henry C. Kapteyn[5], Margaret M. Murnane[5], Robert Streubel[9] and Jianwei Miao[1*]

[1]Department of Physics & Astronomy and California NanoSystems Institute, University of California, Los Angeles, CA 90095, USA and STROBE Science and Technology Center
[2]School of Physical Science and Technology, Soochow University, Suzhou 215006, China
[3]Department of Physics, University of Basel, Klingelbergstrasse 82, CH-4056 Basel, Switzerland
[4]Department of Mathematics, University of California, Los Angeles, CA 90095, USA and STROBE Science and Technology Center
[5]JILA and Department of Physics, University of Colorado and NIST, 440 UCB, Boulder, Colorado 80309, USA and STROBE Science and Technology Center
[6]Polymer Science and Engineering Department University of Massachusetts, Amherst, Massachusetts 01003, USA
[7]Jeremiah Horrocks Institute for Mathematics, Physics and Astronomy, University of Central Lancashire, Preston, PR1 2HE, U.K.
[8]Advanced Light Source, Lawrence Berkeley National Laboratory, Berkeley, CA 94720, USA
[9]Department of Physics & Astronomy, University of Nebraska-Lincoln, Lincoln, NE 68588, USA

## Supplemental Text

**Sample preparation**

The $Fe_3O_4$ nanoparticles (NPs) with a mean diameter of (29.6 ± 2.8) nm [1,3] were dispersed in water. The water solubility of NPs was provided by an amphiphilic polymer coating with a carboxylic acid group and a zeta potential ranging from −35 to −15 mV. The overall thickness of the organic layers was ~4 nm. The inorganic magnetic core was ~22 nm in diameter. A concentration of 1 g/L was used for $Fe_3O_4$ NPs dispersions and the pH of the dispersions was 8, which was adjusted using 1.0 M HCl and measured with a pH meter (Mettler Toledo Electrode Kit EL20 Edu Ph Benc). The $Fe_3O_4$ NPs dispersions with a volume of 5μL was dropped on 500 μm window (10 nm thick Si3N4 and 200 μm thick SiO2 substrate), shown in Fig. S1.

**Data acquisition**

The experiment was conducted at the COSMIC beam line at the Advanced Light Source (Berkeley, CA). An elliptical polarization undulator generated left- and right-circularly polarized x-rays with an energy tuned to the Fe $L_3$ edge (713 eV) and enabled a differential contrast enhancement of the magnetic signal relying on x-ray magnetic circular dichroism [45,46]. The polarized beam was focused onto the sample by a Fresnel zone plate with an outer width of 45 nm. Switching between left- and right-circularly polarized x-rays, tilt series at two in-plane rotation angles (0° and 120°) were recorded (Table. S1). At 0° in-plane rotation, the sample was tilted from -60° to +55° in approximately 5° increments around the *x*-axis. At 120° in-plane angle (rotate 120° around the *z*-axis), the sample was tilted from -50° to +20° in roughly 6° increments. At each tilt angle, the focused beam was raster-scanned across the sample in a 40 nm step size. Diffraction patterns were collected using both left- and right-circularly polarized x-rays. A charge-coupled device camera was used to record the diffraction patterns at each scan position. The full data contains 213,996 diffraction patterns.

**Ptychographic reconstructions**

The real-space images were reconstructed from the diffraction patterns using the extended ptychographic iterative engine (ePIE) [33], in which phase unwrapping and denoising were applied during each ePIE iteration [47]. For high-tilt-angle data, the sample was divided into 15 strip regions, whose probes were individually updated due to defocus-induced blurring effects. Then, all 70 projections were converted to optical density projections [48]. Further background subtraction was performed by numerically evaluating Laplace's equation, using the region exterior to the sample as the boundary condition [35]. The alignment of each pair of oppositely-polarized optical density projections was performed by employing feature-based registration. According to the x-ray magnetic circular dichroism [45,46], the scattering factor contains both electronic and magnetic terms,

$$f = f^c \pm if^m \hat{\mathbf{z}} \cdot \mathbf{S}(\mathbf{r}) \qquad (S1)$$

where $f^c$ and $f^m$ are the charge and magnetic scattering factor, respectively, $\hat{\mathbf{z}}$ is the normal vector along the *z*-axis, $\mathbf{S}(\mathbf{r})$ is the magnetization field at $\mathbf{r} = (x, y, z)$, and '$\pm$' represents left- and right-circularly polarization. The complex-valued ptychography reconstruction is proportional to the scattering factor $f$; thus, the magnetic signal of the material can be retrieved from the difference of the oppositely-polarized optical density. Based on Eq. (S1), 35 projections for 3D scalar and vector reconstructions were computed from the sum and difference of the paired optical density projections, respectively. The common-line and center-of-mass methods were used to align the projections [34].

**3D scalar reconstruction**

The REal Space Iterative REconstruction (RESIRE) algorithm was used for the 3D scalar reconstruction [36]. We first performed two individual 3D reconstructions for the tilt series at the 0° and 120° in-plane angles, from which a transformation matrix was calculated to relate the angles between two tilt series. We then merged two tilt series under the same coordinate system and used

RESIRE to refine the angles. The angle refinement was realized by minimizing the sum of squared errors $\varepsilon(O)$ between experimental and computed projections by,

$$min_O \varepsilon(O) = \sum_{i=1}^{N} \|\Pi_i(O) - b_i\|^2 \qquad (S2)$$

where $O$ is the 3D structure to be reconstructed, $\Pi_i$ is projection operator at Euler angle $\{\varphi_i, \theta_i, \psi_i\}$, $N$ is the number of projections and $b_i$ is the experimental projections. The detailed angular refinement procedure is described elsewhere [36]. After angular refinement, RESIRE was used to obtain the final 3D scalar reconstruction of the sample.

**3D vector reconstruction**

From the final scalar reconstruction, a transformation matrix was computed and applied to the projections for 3D vector reconstruction, which was modeled as a least squares optimization problem,

$$min_{O_1,O_2,O_3} f(O_1, O_2, O_3) = \sum_{i=1}^{N} \|\Pi_i(\alpha_i O_1 + \beta_i O_2 + \gamma_i O_3) - V_i\|^2 \qquad (S3)$$

where $O_1, O_2, O_3$ are the three component of the magnetization vector field to be reconstructed, $V_i$ is the experimental vector projections, $\alpha_i = sin\theta_i cos\varphi_i$, $\beta_i = sin\theta_i sin\varphi_i$ and $\gamma_i = cos\theta_i$. This least square optimization problem was solved directly by gradient descent [25]. To improve the vector reconstruction from a limited number of projections, a 3D support derived from the scalar reconstruction was enforced in the iterative process. A detailed description of the 3D vector reconstruction algorithm can be found elsewhere [26].

**Data analysis**

From the 3D scalar reconstruction, we traced the 3D coordinates of the center of each nanoparticle by using a polynomial fitting method [35,49]. The structural order of the sample was quantified by calculating the radial distribution function (RDF) with the traced center positions. The distance of all particle pairs was calculated and binned into a histogram normalized by the volume of the corresponding spherical shell [48]. The RDF was scaled in order to approach one for infinity. The macrospin for each nanoparticle was calculated by integrating the reconstructed magnetization field inside each nanoparticle. The sample was segmented into the monolayer, bilayer, and trilayer to determine the relationship between the number of the layer and the magnetic order.

**Vortex density**

In the main text, we have introduced the vortex density on a discrete lattice. To gain an intuition, we sketch a vortex [Fig. S4(a)] and antivortex [Fig. S4(b)] configuration, where the vectors rotate counter clock-wisely in Fig. S4(a), but clock-wisely in Fig. S4(b) as we travel counter clock-wisely along the circle. We remark that $\hat{\mathbf{z}} \cdot (\mathbf{S}_l \times \mathbf{S}_{l'})$ captures this vector chirality. In Fig. S4(a), $\hat{\mathbf{z}} \cdot (\mathbf{S}_A \times \mathbf{S}_B) > 0$ corresponds to positive winding. In Fig. S4(b), $\hat{\mathbf{z}} \cdot (\mathbf{S}_A \times \mathbf{S}_B) < 0$ stands for negative winding. Figure S2(a) and (b) show the vortex density map by calculating the barycenter of each pyramid by 3D Delaunay triangulation. The vortex density value calculated by Eq. (3) in the main text was assigned to the corresponding center coordinate of the three neighboring nanoparticles. Note that the maximal value of the vortex number associated with a triangular region is attained when the angles between any two normalized macrospins are equal to $2\pi/3$, resulting in a value of $3\sqrt{3}/4\pi$.

**Micromagnetic Monte Carlo simulations**

The magnetic nanoparticles were modeled as a macrospin ensemble by providing the experimental 3D coordinates to a micromagnetic Monte Carlo method (MMC) [42] implemented in Boris Spintronics [43]. Trial moves of this MMC include not only spin rotations, but also a change in the magnetization length. The method is parameterized using the Curie-Weiss longitudinal susceptibility and reproduces the same Maxwell-Boltzmann distribution for the magnetization length as the stochastic Landau-Lifshitz-Bloch equation [44,50]. It is particularly useful for computing thermodynamic equilibrium states, including simulation of hysteresis loops. A Curie temperature $T_c$ = 850 K was used with a zero-temperature saturation magnetization of 120 kA/m (~105 kA/m at room temperature). The probability distributions of the magnetization length were retrieved from 10,000 ensembles. Vortex density map, spin-spin correlation and vortex-vortex correlation functions were calculated using the same equations introduced in the main text for analyzing the experimental data.

**Supplemental Figures and Table**

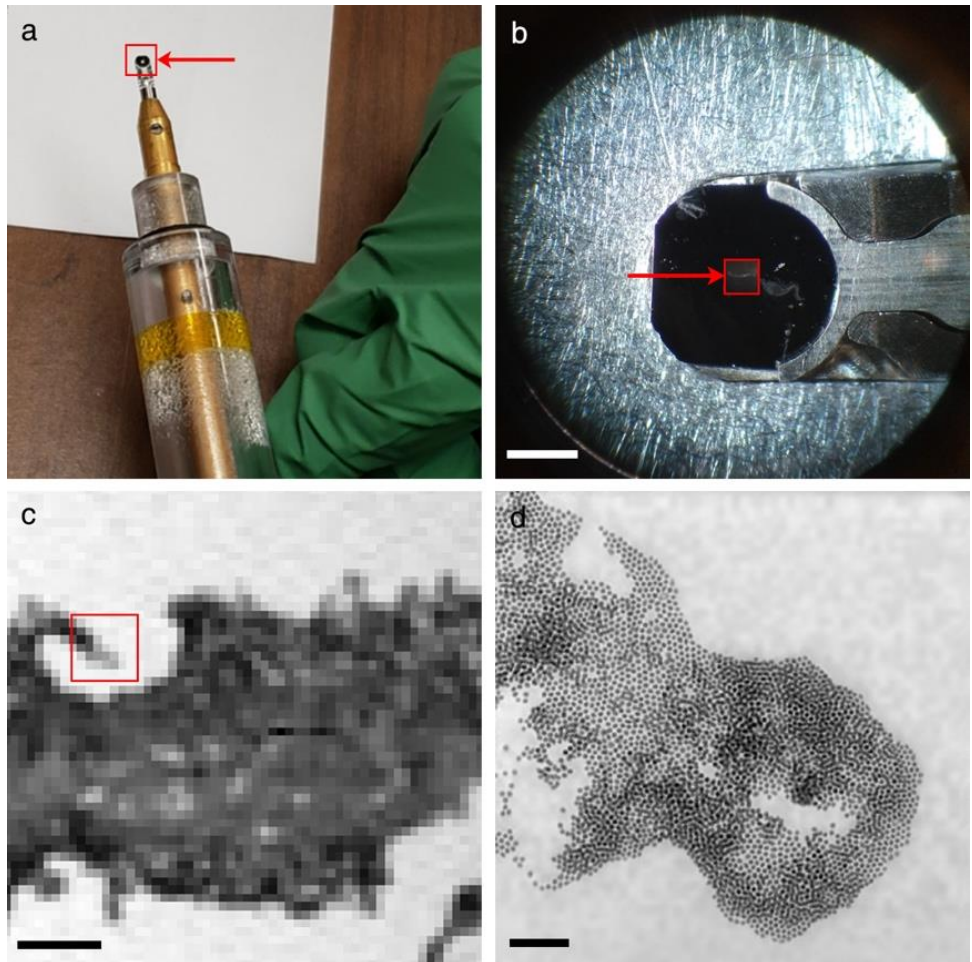

**Fig. S1 |** Sample preparation. (a-b) Picture of self-assembled Fe$_3$O$_4$ nanoparticles on a silicon nitride membrane. (c) Scanning transmission x-ray microscopy images of the sample. (d) Ptychographic reconstruciton of the sqaured region in (c). Scale bars, 1 mm in (b), 10 μm (c) and 300 nm in (d).

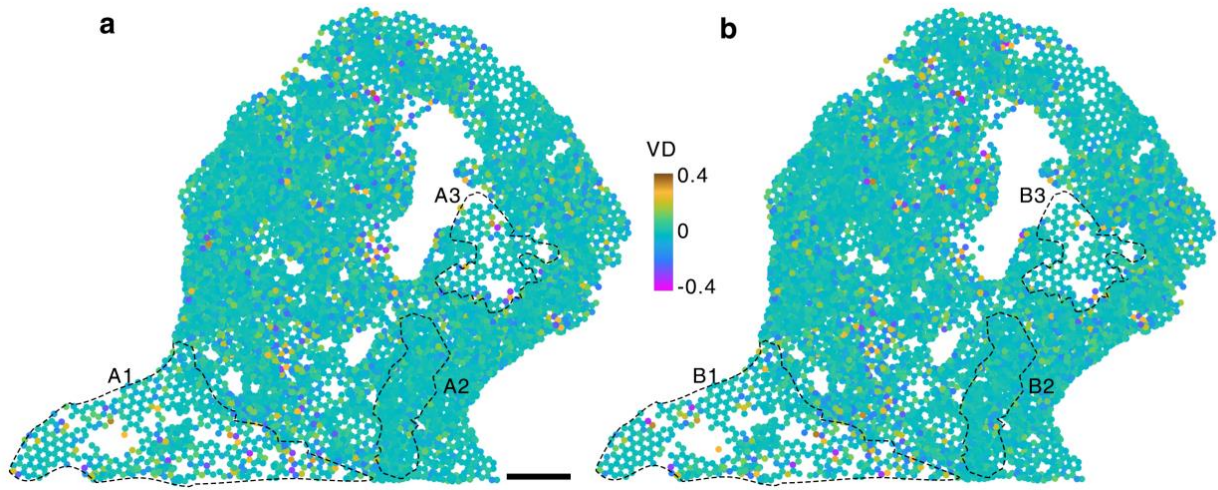

**Fig. S2 |** Vortex density map comparision between experimental and simulated data. The location of macrovortices were obtained from the experimental reconstruction (a) and micromagnetic Monte Carlo simulations (b), where A1/B1, A2/B2, and A3/B3 correspond to A1, A2, and A3 in Fig. 3, respectively.

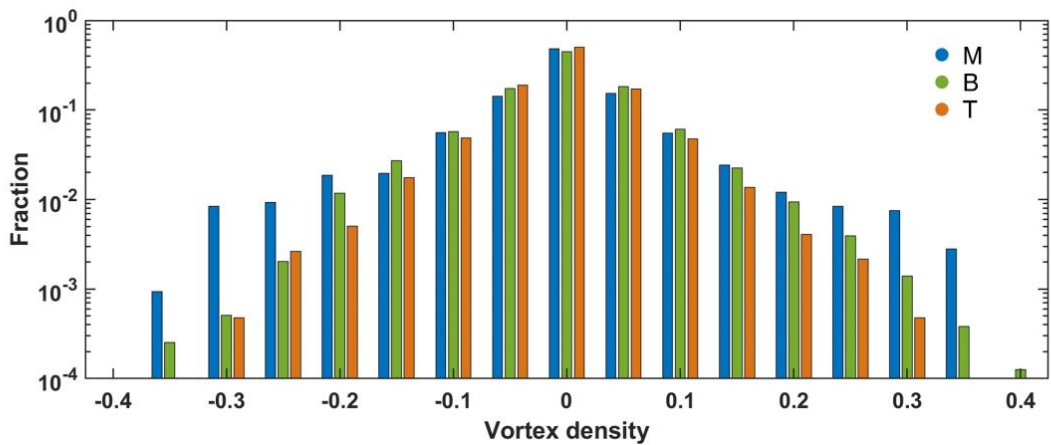

**Fig. S3 |** Histogram of magnetic vortex density in the monolayer (blue), bilayer (green), and trilayer (orange), obtained from Monte Carlo simulations.

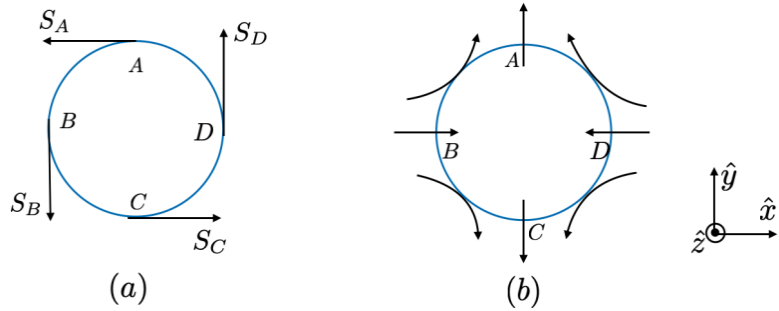

**Fig. S4** | Sketch of a vortex (a) and an antivortex (b) configuration, where the vectors rotate counter clock-wisely in (a), but clock-wisely in (b) as we travel counter clock-wisely along the circle.

| In-plane rotation angle (°) | Tilt angle (°) | | | | | | | | | | | | | | | | | | | | |
|---|---|---|---|---|---|---|---|---|---|---|---|---|---|---|---|---|---|---|---|---|---|
| | #1 | #2 | #3 | #4 | #5 | #6 | #7 | #8 | #9 | #10 | #11 | #12 | #13 | #14 | #15 | #16 | #17 | #18 | #19 | #20 | #21 |
| 0 | -60 | -55 | -50 | -45 | -40 | -30 | -20 | -15 | -10 | -5 | 0 | 5 | 10 | 15 | 20 | 30 | 35 | 40 | 45 | 50 | 55 |
| 120 | -50 | -44 | -38 | -32 | -26 | -20 | -14 | -8 | -2 | 0 | 2 | 8 | 14 | 20 | | | | | | | |

**Table S1** | Tilt angles in two in-plane rotations.